\documentclass[conference]{IEEEtran}
\usepackage{xspace}
\usepackage{url}
\usepackage[cmex10]{amsmath}
\usepackage{bbm}
\usepackage{graphicx}
\usepackage{epstopdf}
\usepackage{paralist}
\usepackage[normalem]{ulem}
\usepackage{fancyref} 
\usepackage{amsmath}
\usepackage{amssymb}
\usepackage[stretch=16,shrink=16,step=4]{microtype}

\usepackage[inline]{enumitem}

\usepackage{xcolor}
\definecolor{links}{rgb}{0.3,0,0}   
\definecolor{urls}{rgb}{0,0,0.8}    
\definecolor{cites}{rgb}{0,0,0.6}   
\usepackage[colorlinks,hyperindex,linkcolor=links,citecolor=cites,urlcolor=urls]{hyperref} 
\usepackage{dsfont}
\usepackage{footnote}
\usepackage{balance}
\usepackage{float}
\usepackage{cite}
\usepackage{wasysym}
\usepackage{amssymb}

\usepackage{subcaption}
\usepackage{tabularx,booktabs}
\newcolumntype{L}[1]{>{\raggedright\let\newline\\\arraybackslash\hspace{0pt}}m{#1}}
\newcolumntype{C}[1]{>{\centering\let\newline\\\arraybackslash\hspace{0pt}}m{#1}}
\newcolumntype{R}[1]{>{\raggedleft\let\newline\\\arraybackslash\hspace{0pt}}m{#1}}
\usepackage{boldline}

\usepackage{vmr-symbols-vecbold}
\usepackage{standard-macros}
\usepackage{mathbbol}

\usepackage{tikz}
\usetikzlibrary{patterns}
\usetikzlibrary{shapes,arrows,positioning}
\usetikzlibrary{calc}
\usetikzlibrary{fit}
\usetikzlibrary{plotmarks}
\tikzset{every picture/.style={font issue=\scriptsize, >=stealth},font issue/.style={execute at begin picture={#1\selectfont}}}
\tikzset{three sided left/.style={
        draw=none,
        xshift=\pgflinewidth,
        append after command={
            [shorten <= -0.5\pgflinewidth]
            ([shift={(-1.5\pgflinewidth,-0.5\pgflinewidth)}]\tikzlastnode.north east) edge ([shift={( 0.5\pgflinewidth,-0.5\pgflinewidth)}]\tikzlastnode.north west) 
            ([shift={( 0.5\pgflinewidth,-0.5\pgflinewidth)}]\tikzlastnode.north west) edge ([shift={( 0.5\pgflinewidth,+0.5\pgflinewidth)}]\tikzlastnode.south west)            
            ([shift={( 0.5\pgflinewidth,+0.5\pgflinewidth)}]\tikzlastnode.south west) edge ([shift={(-1.0\pgflinewidth,+0.5\pgflinewidth)}]\tikzlastnode.south east)
        }}}
        
\tikzset{three sided right/.style={
        draw=none,
        xshift=-\pgflinewidth,
        append after command={
            [shorten <= -0.5\pgflinewidth]
            ([shift={( 1.5\pgflinewidth,-0.5\pgflinewidth)}]\tikzlastnode.north west) edge ([shift={(-0.5\pgflinewidth,-0.5\pgflinewidth)}]\tikzlastnode.north east) 
            ([shift={(-0.5\pgflinewidth,-0.5\pgflinewidth)}]\tikzlastnode.north east) edge ([shift={(-0.5\pgflinewidth,+0.5\pgflinewidth)}]\tikzlastnode.south east)            
            ([shift={(-0.5\pgflinewidth,+0.5\pgflinewidth)}]\tikzlastnode.south east) edge ([shift={( 1.0\pgflinewidth,+0.5\pgflinewidth)}]\tikzlastnode.south west)
        }}}

\usepackage{pgfplots}
\usepgfplotslibrary{fillbetween}
\pgfplotsset{
  compat=newest, 
  width=\columnwidth,    
  height=0.8\columnwidth,   
  plot coordinates/math parser=false,
  standard/.style={
    axis equal,
    axis line style=help lines,
    axis x line=center,
    axis y line=center,
    axis z line=center},
    grid style={dashed,gray},
    minor grid style={dotted,gray},
    major grid style={dotted,gray},
    ylabel absolute, ylabel style={yshift=-0.4cm},
    xlabel absolute, xlabel style={yshift=0.25cm}
}

\makeatletter
 \pgfdeclarepatternformonly[\tikz@pattern@color,\LineSpace,\LineWidth]{my horizontal lines}%
    {\pgfpointorigin}{\pgfqpoint{100pt}{1pt}}{\pgfqpoint{100pt}{\LineSpace}}%
    {
        \pgfsetcolor{\tikz@pattern@color}
        \pgfsetlinewidth{\LineWidth}
        \pgfpathmoveto{\pgfqpoint{0pt}{0.5pt}}
        \pgfpathlineto{\pgfqpoint{100pt}{0.5pt}}
        \pgfusepath{stroke}
    }
 
 \pgfdeclarepatternformonly[\tikz@pattern@color,\LineSpace,\LineWidth]{my vertical lines}%
    {\pgfpointorigin}{\pgfqpoint{1pt}{100pt}}{\pgfqpoint{\LineSpace}{100pt}}%
    {
        \pgfsetcolor{\tikz@pattern@color}
        \pgfsetlinewidth{\LineWidth}
        \pgfpathmoveto{\pgfqpoint{0.5pt}{0pt}}
        \pgfpathlineto{\pgfqpoint{0.5pt}{100pt}}
        \pgfusepath{stroke}
    } 
 
 \pgfdeclarepatternformonly[\tikz@pattern@color,\LineSpace,\LineWidth]{my grid}%
    {\pgfqpoint{-1pt}{-1pt}}{\pgfqpoint{\LineSpace}{\LineSpace}}
    {\pgfqpoint{\LineSpace}{\LineSpace}}%
    {
        \pgfsetcolor{\tikz@pattern@color}
        \pgfsetlinewidth{\LineWidth}
        \pgfpathmoveto{\pgfqpoint{0pt}{0pt}}
        \pgfpathlineto{\pgfqpoint{0pt}{\LineSpace + 0.1pt}}
        \pgfpathmoveto{\pgfqpoint{0pt}{0pt}}
        \pgfpathlineto{\pgfqpoint{\LineSpace + 0.1pt}{0pt}}
        \pgfusepath{stroke}
    }
 
 \pgfdeclarepatternformonly[\tikz@pattern@color,\LineSpace,\LineWidth]{my north east lines}
    {\pgfqpoint{-\LineWidth}{-\LineWidth}}{\pgfqpoint{\LineSpace}{\LineSpace}}
    {\pgfqpoint{\LineSpace}{\LineSpace}}%
    {
        \pgfsetcolor{\tikz@pattern@color}
        \pgfsetlinewidth{\LineWidth}
        \pgfpathmoveto{\pgfqpoint{-\LineWidth}{-\LineWidth}}
        \pgfpathlineto{\pgfqpoint{\LineSpace + 0.1pt}{\LineSpace + 0.1pt}}
        \pgfusepath{stroke}
    }
 
 \pgfdeclarepatternformonly[\tikz@pattern@color,\LineSpace,\LineWidth]{my north west lines}
    {\pgfqpoint{-\LineWidth}{-\LineWidth}}{\pgfqpoint{\LineSpace}{\LineSpace}}
    {\pgfqpoint{\LineSpace}{\LineSpace}}%
    {
        \pgfsetcolor{\tikz@pattern@color}
        \pgfsetlinewidth{\LineWidth}
        \pgfpathmoveto{\pgfqpoint{-\LineWidth}{\LineSpace}}
        \pgfpathlineto{\pgfqpoint{\LineSpace + 0.1pt}{-\LineWidth}}
        \pgfusepath{stroke}
    }
 
 \pgfdeclarepatternformonly[\tikz@pattern@color,\LineSpace,\LineWidth]{my crosshatch}%
    {\pgfqpoint{-1pt}{-1pt}}{\pgfqpoint{\LineSpace}{\LineSpace}}
    {\pgfqpoint{\LineSpace}{\LineSpace}}%
    {
        \pgfsetcolor{\tikz@pattern@color}
        \pgfsetlinewidth{\LineWidth}
        \pgfpathmoveto{\pgfqpoint{\LineSpace + 0.1pt}{0pt}}
        \pgfpathlineto{\pgfqpoint{0pt}{\LineSpace + 0.1pt}}
        \pgfpathmoveto{\pgfqpoint{0pt}{0pt}}
        \pgfpathlineto{\pgfqpoint{\LineSpace + 0.1pt}{\LineSpace + 0.1pt}}
        \pgfusepath{stroke}
    }
 
 \pgfdeclarepatternformonly[\tikz@pattern@color,\LineSpace,\PointSize]{my dots}%
    {\pgfqpoint{-\LineSpace*0.25}{-\LineSpace*0.25}}
    {\pgfqpoint{\LineSpace*0.25}{\LineSpace*0.25}}
    {\pgfqpoint{\LineSpace*0.75}{\LineSpace*0.75}}%
    {
        \pgfsetcolor{\tikz@pattern@color}
        \pgfpathcircle{\pgfqpoint{0pt}{0pt}}{\PointSize}
        \pgfusepath{fill}
    }
\makeatother
 
\newdimen\LineSpace
\newdimen\PointSize
\newdimen\LineWidth
\tikzset{
    line space/.code={\LineSpace=#1},
    line space=3pt
}
\tikzset{
    point size/.code={\PointSize=#1},
    point size=.5pt
}
\tikzset{
    pattern line width/.code={\LineWidth=#1},
    pattern line width=.4pt
}

\DeclareSymbolFontAlphabet{\amsmathbb}{AMSb}%

\newcommand{\lro}[1]{\lefto({#1}\right)}																
\newcommand{\lrbo}[1]{\lefto \lbrace {#1} \right \rbrace}															
\newcommand{\lrho}[1]{\lefto [ {#1} \right ]}																				

\newcommand{\lr}[1]{\left({#1}\right)}																
\newcommand{\lrb}[1]{\left \lbrace {#1} \right \rbrace}															

\safemath{\dopplerspread}{B_D}																								
\safemath{\delayspread}{T_D}																									
\safemath{\nc}{n\sub{c}}																										
\safemath{\nf}{n\sub{f}}																										
\safemath{\efa}{\epsilon\sub{FA}}
\safemath{\emd}{\epsilon\sub{MD}}
\safemath{\ie}{\epsilon\sub{IE}}
\safemath{\e}{\epsilon}
\safemath{\ed}{\epsilon\sub{D}}
\safemath{\ex}{\epsilon\sub{EE}}

\safemath{\ad}{\alpha\sub{d}}
\safemath{\afa}{\alpha\sub{fa}}
\safemath{\amd}{\alpha\sub{md}}

\safemath{\Je}{\widehat{J}}
\safemath{\nd}{n\sub{d}}																										
\safemath{\ntx}{n\sub{t}} 																											
\safemath{\nrx}{n\sub{r}}																											
\safemath{\ntxt}{\tilde{n\sub{t}}}																											
\safemath{\cb}{\ensuremath{L}} 																								
\safemath{\cl}{\ensuremath{n}} 																								
\safemath{\txanto}{{\ensuremath{\tilde{m}_t}}} 																		
\safemath{\cs}{M} 																														
\safemath{\idPustm}{\ensuremath{S_{k}}}
\safemath{\error}{\ensuremath{\epsilon}} 																				
\safemath{\eexp}{\ensuremath{\mathcal{E}}} 																			
\safemath{\nsubc}{n\sub{s}}			 																						
\safemath{\nofdm}{n\sub{o}} 																									
\safemath{\bc}{\ensuremath{B_c}} 																							
\safemath{\ts}{\ensuremath{T_s}} 																							
\safemath{\nrb}{\ensuremath{n_{rb}}} 																						
\safemath{\nres}{\ell}
\safemath{\maxk}{M^*\lr{\nres, \nsubc, \nofdm, \epsilon, \rho}}
\safemath{\Rmax}{R^*}
\safemath{\Emin}{E\sub{b}^*/N_0}
\safemath{\Eminf}{\frac{E\sub{b}^*}{N_0}}
\safemath{\np}{\ensuremath{n\sub{p}}}
\safemath{\code}{\ensuremath{\mathcal{C}}}
\safemath{\err}{\ensuremath{\epsilon}}
\safemath{\rp}{\ensuremath{\rho\sub{p}}}
\safemath{\rd}{\ensuremath{\rho\sub{d}}}
\safemath{\cohtime}{\ensuremath{T\sub{c}}}
\safemath{\cohbw}{\ensuremath{B\sub{c}}}
\safemath{\nmax}{\ensuremath{\ell\sub{max}}}

\safemath{\yp}{\ensuremath{\randmatY^{(\text{p})}}}
\safemath{\yd}{\ensuremath{\randmatY^{(\text{d})}}}

\safemath{\xp}{\ensuremath{\matX^{(\text{p})}}}
\safemath{\xd}{\ensuremath{\randmatX^{(\text{d})}}}
\safemath{\xpbar}{\ensuremath{\overline{\matX}^{(\text{p})}}}
\safemath{\xdbar}{\ensuremath{\overline{\randmatX}^{(\text{d})}}}

\safemath{\xdv}{\ensuremath{\randvecx^{(\text{d})}}}
\safemath{\xdbarv}{\ensuremath{\overline{\randvecx}^{(\text{d})}}}
\safemath{\ydv}{\ensuremath{\randvecy^{(\text{d})}}}

\safemath{\xdr}{\ensuremath{\matX^{(\text{d})}}}

\safemath{\mI}{\ensuremath{i\lro{\randvecy ; \randvecx}}} 				

\safemath{\randveca}{\bm{A}}
\safemath{\randvecb}{\bm{B}}
\safemath{\randvecc}{\bm{C}}
\safemath{\randvecd}{\bm{D}}
\safemath{\randvece}{\bm{E}}
\safemath{\randvecf}{\bm{F}}
\safemath{\randvecg}{\bm{G}}
\safemath{\randvech}{\bm{H}}
\safemath{\randveci}{\bm{I}}
\safemath{\randvecj}{\bm{J}}
\safemath{\randveck}{\bm{K}}
\safemath{\randvecl}{\bm{L}}
\safemath{\randvecm}{\bm{M}}
\safemath{\randvecn}{\bm{N}}
\safemath{\randveco}{\bm{O}}
\safemath{\randvecp}{\bm{P}}
\safemath{\randvecq}{\bm{Q}}
\safemath{\randvecr}{\bm{R}}
\safemath{\randvecs}{\bm{S}}
\safemath{\randvect}{\bm{T}}
\safemath{\randvecu}{\bm{U}}
\safemath{\randvecv}{\bm{V}}
\safemath{\randvecw}{\bm{W}}
\safemath{\randvecx}{\bm{X}}
\safemath{\randvecy}{\bm{Y}}
\safemath{\randvecz}{\bm{Z}}
\safemath{\randvecphi}{\bm{\Phi}}

\safemath{\randmatA}{\amsmathbb{A}}
\safemath{\randmatB}{\amsmathbb{B}}
\safemath{\randmatC}{\amsmathbb{C}}
\safemath{\randmatD}{\amsmathbb{D}}
\safemath{\randmatE}{\amsmathbb{E}}
\safemath{\randmatF}{\amsmathbb{F}}
\safemath{\randmatG}{\amsmathbb{G}}
\safemath{\randmatH}{\amsmathbb{H}}
\safemath{\randmatI}{\amsmathbb{I}}
\safemath{\randmatJ}{\amsmathbb{J}}
\safemath{\randmatK}{\amsmathbb{K}}
\safemath{\randmatL}{\amsmathbb{L}}
\safemath{\randmatM}{\amsmathbb{M}}
\safemath{\randmatN}{\amsmathbb{N}}
\safemath{\randmatO}{\amsmathbb{O}}
\safemath{\randmatP}{\amsmathbb{P}}
\safemath{\randmatQ}{\amsmathbb{Q}}
\safemath{\randmatR}{\amsmathbb{R}}
\safemath{\randmatS}{\amsmathbb{S}}
\safemath{\randmatT}{\amsmathbb{T}}
\safemath{\randmatU}{\amsmathbb{U}}
\safemath{\randmatV}{\amsmathbb{V}}
\safemath{\randmatW}{\amsmathbb{W}}
\safemath{\randmatX}{\amsmathbb{X}}
\safemath{\randmatY}{\amsmathbb{Y}}
\safemath{\randmatZ}{\amsmathbb{Z}}
\safemath{\randmatSigma}{\mathbb{\Sigma}}
\safemath{\randmatPhi}{\mathbb{\Phi}}
\safemath{\randmatLambda}{\mathbb{\Lambda}}

\safemath{\matSigma}{\bm{\Sigma}}
\safemath{\matPhi}{\bm{\Phi}}
\safemath{\matLambda}{\bm{\Lambda}}

\safemath{\T}{T}

\usepackage{glossaries}
\newglossaryentry{sw}
{
  name={SW},
  description={synchronization word},
  first={\glsentrydesc{sw} (\glsentrytext{sw})},
  plural={SW's},
  descriptionplural={synchronization words},
  firstplural={\glsentrydescplural{sw} (\glsentryplural{sw})}
}

\newglossaryentry{uep}
{
  name={UEP},
  description={unequal error protection},
  first={\glsentrydesc{uep} (\glsentrytext{uep})}
}

\newglossaryentry{md}
{
  name={MD},
  description={misdetection},
  first={\glsentrydesc{md} (\glsentrytext{md})}
}

\newglossaryentry{fa}
{
  name={FA},
  description={false alarm},
  first={\glsentrydesc{fa} (\glsentrytext{fa})}
}

\newglossaryentry{ie}
{
  name={IE},
  description={inclusive error},
  first={\glsentrydesc{ie} (\glsentrytext{ie})}
}

\newglossaryentry{biawgn}
{
  name={BI-AWGN},
  description={binary-input additive white Gaussian noise},
  first={\glsentrydesc{biawgn} (\glsentrytext{biawgn})}
}
\newglossaryentry{bsc}
{
  name={BSC},
  description={binary-symmetric channel},
  first={\glsentrydesc{bsc} (\glsentrytext{bsc})}
}

\newglossaryentry{bec}
{
  name={BEC},
  description={binary-erasure channel},
  first={\glsentrydesc{bec} (\glsentrytext{bec})}
}
\newglossaryentry{awgn}
{
  name={AWGN},
  description={additive white Gaussian noise},
  first={\glsentrydesc{awgn} (\glsentrytext{awgn})}
}
\newglossaryentry{3gpp}
{
  name={3GPP},
  description={3rd generation partnership project},
  first={\glsentrydesc{3gpp} (\glsentrytext{3gpp})}
}

\newglossaryentry{dl}
{
  name={DL},
  description={downlink},
  first={\glsentrydesc{dl} (\glsentrytext{dl})}
}

\newglossaryentry{LED}
{
  name={LED},
  description={light emitting diode},
  first={\glsentrydesc{LED} (\glsentrytext{LED})},
  plural={LEDs},
  descriptionplural={light emitting diodes},
  firstplural={\glsentrydescplural{LED} (\glsentryplural{LED})}
}

\newglossaryentry{lte}
{
  name={LTE},
  description={long term evolution},
  first={\glsentrydesc{lte} (\glsentrytext{lte})}
}
\newglossaryentry{ofdm}
{
  name={OFDM},
  description={orthogonal frequency-division multiplexing},
  first={\glsentrydesc{ofdm} (\glsentrytext{ofdm})}
}
\newglossaryentry{ue}
{
  name={UE},
  description={user equipment},
  first={\glsentrydesc{ue} (\glsentrytext{ue})},
  plural={UE's},
  descriptionplural={user equipments},
  firstplural={\glsentrydescplural{ue} (\glsentryplural{ue})}
}
\newglossaryentry{ul}
{
  name={UL},
  description={uplink},
  first={\glsentrydesc{ul} (\glsentrytext{ul})}
}
\newglossaryentry{rb}
{
  name={RB},
  description={resource block},
  first={\glsentrydesc{rb} (\glsentrytext{rb})},
  plural={RBs},
  descriptionplural={resource blocks},
  firstplural={\glsentrydescplural{rb} (\glsentryplural{rb})}
}
\newglossaryentry{cu}
{
  name={CU},
  description={channel use},
  first={\glsentrydesc{cu} (\glsentrytext{cu})},
  plural={CUs},
  descriptionplural={channel uses},
  firstplural={\glsentrydescplural{cu} (\glsentryplural{cu})}
}
\newglossaryentry{tti}
{
  name={TTI},
  description={transmission time interval},
  first={\glsentrydesc{tti} (\glsentrytext{tti})},
  plural={TTI's},
  descriptionplural={transmission time intervals},
  firstplural={\glsentrydescplural{tti} (\glsentryplural{tti})}
}
\newglossaryentry{iid}
{
  name={i.i.d.},
  description={identical and independently distributed},
  first={\glsentrydesc{iid} (\glsentrytext{iid})}
}
\newglossaryentry{psd}
{
  name={PSD},
  description={power spectral density},
  first={\glsentrydesc{psd} (\glsentrytext{psd})}
}

\newglossaryentry{mimo}
{
  name={MIMO},
  description={multiple-input multiple-output},
  first={\glsentrydesc{mimo} (\glsentrytext{mimo})}
}

\newglossaryentry{bler}
{
  name={BLER},
  description={block error probability},
  first={\glsentrydesc{bler} (\glsentrytext{bler})}
}
\newglossaryentry{mtc}
{
  name={MTC},
  description={machine-type communication},
  first={\glsentrydesc{mtc} (\glsentrytext{mtc})}
}
\newglossaryentry{csi}
{
  name={CSI},
  description={channel state information},
  first={\glsentrydesc{csi} (\glsentrytext{csi})}
}
\newglossaryentry{dvbt}
{
  name={DVB-T},
  description={terrestrial digital video broadcasting},
  first={\glsentrydesc{dvbt} (\glsentrytext{dvbt})}
}
\newglossaryentry{pat}
{
  name={PAT},
  description={pilot-assisted transmission},
  first={\glsentrydesc{pat} (\glsentrytext{pat})}
}
\newglossaryentry{ml}
{
  name={ML},
  description={maximum likelihood},
  first={\glsentrydesc{ml} (\glsentrytext{ml})}
}
\newglossaryentry{dt}
{
  name={DT},
  description={dependence-testing},
  first={\glsentrydesc{dt} (\glsentrytext{dt})}
}

\newglossaryentry{mc}
{
  name={MC},
  description={meta-converse},
  first={\glsentrydesc{mc} (\glsentrytext{mc})}
}
\newglossaryentry{ustm}
{
  name={USTM},
  description={unitary space-time modulation},
  first={\glsentrydesc{ustm} (\glsentrytext{ustm})}
}
\newglossaryentry{siso}
{
  name={SISO},
  description={single-input single-output},
  first={\glsentrydesc{siso} (\glsentrytext{siso})}
}
\newglossaryentry{snn}
{
  name={SNN},
  description={scaled nearest-neighbor},
  first={\glsentrydesc{snn} (\glsentrytext{snn})}
}
\newglossaryentry{rcus}
{
  name={RCUs},
  description={random-coding union bound with parameter $s$},
  first={\glsentrydesc{rcus} (\glsentrytext{rcus})}
}
\newglossaryentry{osd}
{
  name={OSD},
  description={ordered statistics decoding},
  first={\glsentrydesc{osd} (\glsentrytext{osd})}
}
\newglossaryentry{llr}
{
  name={LLR},
  description={log-likelihood ratio},
  first={\glsentrydesc{llr} (\glsentrytext{llr})}
   plural={LLR's},
  descriptionplural={log-likelihood ratios},
  firstplural={\glsentrydescplural{llr} (\glsentryplural{llr})}
}
\newglossaryentry{qpsk}
{
  name={QPSK},
  description={quaternary phase shift keying},
  first={\glsentrydesc{qpsk} (\glsentrytext{qpsk})}
}
\newglossaryentry{nlos}
{
  name={NLOS},
  description={non-line-of-sight},
  first={\glsentrydesc{nlos} (\glsentrytext{nlos})}
}
\newglossaryentry{los}
{
  name={LOS},
  description={line-of-sight},
  first={\glsentrydesc{los} (\glsentrytext{los})}
}
\newglossaryentry{cgf}
{
  name={CGF},
  description={cumulant-generating function},
  first={\glsentrydesc{cgf} (\glsentrytext{cgf})},
   plural={CGF's},
  descriptionplural={cumulant-generating functions},
  firstplural={\glsentrydescplural{cgf} (\glsentryplural{cgf})}
}
\newglossaryentry{pdf}
{
  name={PDF},
  description={probability density function},
  first={\glsentrydesc{pdf} (\glsentrytext{pdf})},
  plural={PDFs},
  descriptionplural={probability density functions},
  firstplural={\glsentrydescplural{pdf} (\glsentryplural{pdf})}
}
\newglossaryentry{gmi}
{
  name={GMI},
  description={generalized mutual information},
  first={\glsentrydesc{gmi} (\glsentrytext{gmi})}
}

\newglossaryentry{arq}
{
  name={ARQ},
  description={automatic repeat request},
  first={\glsentrydesc{arq} (\glsentrytext{arq})}
}

\newglossaryentry{harq}
{
  name={HARQ},
  description={hybrid automatic repeat request},
  first={\glsentrydesc{harq} (\glsentrytext{harq})}
}

\newglossaryentry{fbl}
{
  name={FBL-NF},
  description={fixed blocklength with no feedback},
  first={\glsentrydesc{fbl} (\glsentrytext{fbl})}
}
\newglossaryentry{ssnf}
{
  name={SS-NF},
  description={single-shot no-feedback},
  first={\glsentrydesc{ssnf} (\glsentrytext{ssnf})}
}

\newglossaryentry{urllc}
{
  name={URLLC},
  description={ultra-reliable low-latency communications},
  first={\glsentrydesc{urllc} (\glsentrytext{urllc})}
}

\newglossaryentry{vlsf}
{
  name={VLSF},
  description={variable-length stop-feedback},
  first={\glsentrydesc{vlsf} (\glsentrytext{vlsf})}
}

\newglossaryentry{dmt}
{
  name={DMT},
  description={diversity-multiplexing tradeoff},
  first={\glsentrydesc{dmt} (\glsentrytext{dmt})}
}

\newglossaryentry{dmdt}
{
  name={DMDT},
  description={diversity-multiplexing-delay tradeoff},
  first={\glsentrydesc{dmdt} (\glsentrytext{dmdt})}
}

\newglossaryentry{tddt}
{
  name={TDDT},
  description={throughput-diversity-delay tradeoff},
  first={\glsentrydesc{tddt} (\glsentrytext{tddt})}
}

\newglossaryentry{stbc}
{
  name={STBC},
  description={space-time block-codes},
  first={\glsentrydesc{stbc} (\glsentrytext{stbc})},
  plural={STBC's},
  descriptionplural={space-time block-codes},
  firstplural={\glsentrydescplural{stbc} (\glsentryplural{stbc})}
}

\newglossaryentry{harqir}
{
  name={HARQ-IR},
  description={hybrid automatic repeat request with incremental redundancy},
  first={\glsentrydesc{harqir} (\glsentrytext{harqir})}
}

\newglossaryentry{harqcc}
{
  name={HARQ-CC},
  description={hybrid automatic repeat request with chase combining},
  first={\glsentrydesc{harqcc} (\glsentrytext{harqcc})}
}

\newglossaryentry{wp1}
{
  name={w.p.1.},
  description={with probability one},
  first={\glsentrydesc{wp1} (\glsentrytext{wp1})}
}

\newglossaryentry{vlf}
{
  name={VLF},
  description={variable-length feedback},
  first={\glsentrydesc{vlf} (\glsentrytext{vlf})}
}

\newglossaryentry{dmc}
{
  name={DMC},
  description={discrete memoryless channel},
  first={\glsentrydesc{dmc} (\glsentrytext{dmc})},
  plural={DMCs},
  descriptionplural={discrete memoryless channels},
  firstplural={\glsentrydescplural{dmc} (\glsentryplural{dmc})}
}

\usepackage{subcaption}
\usepackage{pgfplots}
\usepgfplotslibrary{groupplots}
\usetikzlibrary{pgfplots.groupplots}

\usepgfplotslibrary{external}

\newcommand\linew{1pt} 
\newcommand{\fwidth}{\textwidth}

\makeatletter
\def\@IEEEinterspaceratioM{0.265}
\def\@IEEEinterspaceMINratioM{0.1651}
\def\@IEEEinterspaceMAXratioM{0.38}

\def\@IEEEinterspaceratioB{0.31}
\def\@IEEEinterspaceMINratioB{0.19}
\def\@IEEEinterspaceMAXratioB{0.38}
\@IEEEtunefonts
\makeatother
\begin{document}
\IEEEoverridecommandlockouts
\title{On Joint Detection and Decoding \\ in Short-Packet Communications}

\author{
\IEEEauthorblockN{Alejandro Lancho, Johan \"Ostman, and Giuseppe Durisi}
\IEEEauthorblockA{
Chalmers University of Technology, Gothenburg, Sweden\\Emails: \{lanchoa, johanos, durisi\}@chalmers.se}
\thanks{This work was supported by the Swedish Research Council under grant 2016-03293 and by the Wallenberg AI, autonomous systems, and software program.}

}

\maketitle

\begin{abstract}
We consider a communication problem in which the receiver must first detect the presence of an information packet and, if detected, decode
the message carried within it.
We present general nonasymptotic upper and lower bounds on the maximum coding rate that depend on the blocklength, the probability of false alarm,
the probability of misdetection, and the packet error probability.
The bounds, which are expressed in terms of binary-hypothesis-testing performance metrics, 
generalize finite-blocklength bounds derived previously for the scenario when a genie informs the receiver
whether a packet is present. 
The bounds apply to detection performed either jointly with decoding on the entire data packet, or separately on a dedicated preamble. 
The results presented in this paper can be used to determine the blocklength values at which the performance of a communication system is limited
by its ability to perform packet detection satisfactorily, and to assess the difference in performance between preamble-based detection,
and joint detection and decoding.  
Numerical results pertaining to the binary-input AWGN channel are provided.
\end{abstract}
\section{Introduction}\label{sec:intro}
The design of short-packet communications---an integral component of delay-sensitive information-exchange protocols---is subject to
nontrivial tradeoffs between rate, delay, and error probability~\cite{polyanskiy10-05a}.
During the last decade, these tradeoffs have been characterized for many channels of interest, including the \gls{awgn} channel~\cite{polyanskiy10-05a}, and Rayleigh and Rician fading channels~\cite{yang14-07c,durisi16-02a,Lancho20,collins19-01,Ostman19-02}.
These works all rely on the assumption that the receiver has correctly decoded the presence of a data packet, and do not account for the
cost of packet detection. 
However, there exists a plethora of practical applications, including, e.g., sensor networks, event-triggered communications, and
random-access protocols, in which the cost of packet detection is not negligible~\cite{Wang10,Bana18-04,Weinberger17-10a}. 

A common approach to perform detection in such systems is to incorporate within the data packet a deterministic preamble, known to the receiver.
In short-packet applications, however, adding  such a preamble may be highly suboptimal, due to the limited size of the data packet.
Naturally, the following question arises: how much can be gained from a design in which detection and decoding are performed jointly over
the entire data packet, without the insertion of a dedicated preamble?

\paragraph*{Prior Art}
Consider a frame-synchronous system, in which each frame may be empty or may contain a data packet,  \pagebreak  and where the receiver is assumed to
have acquired perfect frame synchronization.
The task of the receiver is to detect whether a packet is present in a frame, and if so, to decode it. 
Three types of error events can be defined: 
a \gls{fa}, i.e., the event that the receiver detects the presence of a packet even though the transmitter was idle; 
a \gls{md}, i.e., the event that the receiver erroneously decides that the transmitter was idle; 
an \gls{ie}, i.e., the event that the receiver does not decode correctly a transmitted codeword.

{An error-exponent analysis} of joint detection and decoding was {first} presented in~\cite{Wang10}, where the random-coding error exponents $(E\sub{\gls{fa}}(R), E\sub{\gls{md}}(R), E\sub{\gls{ie}}(R))$ of the aforementioned errors were analyzed for a given rate $R$ over a \gls{dmc}.
Specifically, the region $(E\sub{\gls{fa}}(R), 0, 0)$ was characterized exactly and the region $(E\sub{\gls{fa}}(R), E\sub{\gls{md}}(R), 0)$ was characterized in terms of inner and outer bounds~\cite[Ch.~3--4]{Wang10}.
It was also shown that separate detection and decoding strategies are strictly suboptimal for all rates and that the gap from optimality grows with the rate.
In~\cite{Weinberger14-09a}, the optimal joint detection and decoding rule for a given code and given \gls{fa} and \gls{md} constraints was derived. 
Furthermore, the authors provide an exact characterization of the corresponding $(E\sub{\gls{fa}}(R), E\sub{\gls{md}}(R), E\sub{\gls{ie}}(R))$-region for the case of constant-composition codes.
The results in~\cite{Weinberger14-09a} were further extended in~\cite{Bayram16} to account for nonuniformly distributed message sets. 

The joint detection and decoding problem is related to the \gls{uep} problem, in which messages belong to different classes with different reliability requirements~\cite{Borade09}. 
A nonasymptotic analysis of the \gls{uep} problem was presented in~\cite{Shkel15}, where several results from~\cite{polyanskiy10-05a} are extended to this setting.
In particular, the \gls{dt} achievability bound (maximum error probability)~\cite[Th.~21]{polyanskiy10-05a} and the metaconverse
bound~\cite[Th.~27]{polyanskiy10-05a} are extended to both joint and separate message classification and decoding, and particularized to the \gls{bsc} and the \gls{bec}. 
However, these bounds cannot be applied directly to the detection and decoding problem considered in this paper, 
since the \gls{md} probability cannot be accounted for straightforwardly. 

In summary, although prior art is available for the problem of joint detection and decoding, in none of the existing
works nonasymptotic bounds are computed for channel models of practical interest in wireless communications.
Furthermore, the performance gap between joint detection and decoding and conventional preamble-based detection followed by decoding is not quantified.

\paragraph*{Contributions}

{We present finite-blocklength achievability and converse bounds for joint detection and decoding strategies over general point-to-point channels.}
The bounds are explicit in the Neyman-Pearson $\beta$ function---a key performance metric in binary hypothesis testing. 
Specifically, the achievability bound builds upon the $\beta\beta$ achievability bound presented in~\cite[Th.1]{Yang18},
while the converse bound is based on the $\beta\beta$ converse bound~\cite[Th.~15]{polyanskiy14-01}, a tightening of the metaconverse bound~\cite[Th.~27]{polyanskiy10-05a}. 
We also discuss how to adapt these bounds to the case of detection performed on a dedicated preamble.

For given requirements on the \gls{fa}, \gls{md}, and \gls{ie} probabilities,
we use the bounds to characterize the maximum coding rate as a function of the packet length both for joint detection and decoding and for
detection performed on a dedicated preamble, followed by decoding.
Numerical results for the binary-input \gls{awgn} channel indicate that packet detection, even when performed jointly with
decoding, is the performance bottleneck when the
packets are very short (less than $190$ channel uses), and the SNR is low (e.g., $0\dB$).
In this regime, it turns out useful to consider input distributions for which the binary input symbols are transmitted with different probabilities, since
this is beneficial for packet detection. 
As the packet length grows and/or the SNR increases, packet decoding becomes the performance bottleneck.
Performing detection solely on the basis of a dedicated preamble turns out to be highly suboptimal in the short-packet regime, even when the length of the preamble is optimized.

\paragraph*{Notation}%
Uppercase letters such as $X$ and $\randvecx$ are used to denote scalar random variables and vectors, respectively; their realizations are written in lowercase, e.g., $x$ and $\vecx$.
We denote the expectation operator by $\mathbb{E}[\cdot]$, and, for a given event~$\setE$, we use $P_{X}[\setE]$ to denote the probability
that $\setE$ occurs when the underlying probability measure is $P_{X}$. 
We write $\log(\cdot)$ and $\log_2(\cdot)$ to denote the natural logarithm and the logarithm to the base $2$, respectively, and $\mathbb{1}[\cdot]$ to denote the indicator function.
Finally, consider two probability distributions $P_{X}$ and $Q_{X}$ on a common measurable space $\setX$, and let $P_{Z\given X}:\setX \rightarrow \{0,1\}$ be a randomized
binary test, where $Z=1$ indicates that the test chooses $P_{X}$. 
The Neyman-Pearson $\beta$ function is defined as 
\begin{equation}\label{eq:np-beta}
  \beta_\alpha(P_X,Q_X) \triangleq \min_{P_{Z\given X}: P_X[Z=1] \geq \alpha} Q_X[Z=1]
\end{equation}
where $\alpha \in [0,1]$.
The test achieving~\eqref{eq:np-beta} is the Neyman-Pearson test~\cite{neyman33-01a}, which involves thresholding the log-likelihood ratio $\log
\mathrm{d}P_{X}/\mathrm{d}Q_{X}$, where $\mathrm{d}P_{X}/\mathrm{d}Q_{X}$ stands for the Radon-Nykodim derivative of $P_{X}$ with respect to
$Q_{X}$. 
We shall also use for convenience the following equivalent function
\begin{equation}
  \alpha_\beta(P_X,Q_X) \triangleq \min_{P_{Z\given X}: Q_X[Z=1] \leq \beta} P_X[Z=0]
\end{equation}
where $\beta \in \lrho{0,1}$.
This function turns out to be easier to relate to error probabilities than~\eqref{eq:np-beta}.

\section{System Model}\label{sec:syst_model}
We consider frame-synchronous transmissions where the packet duration is equal to the frame size and encompasses~$n$ channel uses.
The receiver is assumed to be oblivious to the presence of any packet and must, therefore, perform both detection and decoding.
The channel law within each frame is denoted by $P_{\randvecy \given \randvecx}(\vecy \given \vecx)$, where $\vecx \in  \mathcal{X}^n$ and 
 $\vecy \in \mathcal{Y}^n$.
 Throughout, we assume that $\mathcal{X}$ contains a special symbol, which we denote by~$\varnothing$, used to indicate that the transmitter
 is idle.
 We shall use $\boldsymbol{\varnothing}$ to indicate the vector containing $n$ copies of $\varnothing$. 

When the transmitter is active, it encodes a message $W$, uniformly distributed in the set $\{1,\dots,M\}$, to a codeword in the codebook
$\{\vecc_i\}_{i=1}^M \subset \mathcal{X}^{n}$.
To perform joint detection and decoding, the receiver partitions the set $\mathcal{Y}^{n}$ of possible received signals $\vecy$ into  $M+1$
disjoint regions $\mathcal{R}_0, \mathcal{R}_1, \dots, \mathcal{R}_M$ whose union covers $\mathcal{Y}^n$. 
If $\vecy \in \mathcal{R}_m$, where $m = 1, \dots, M$, the decoder returns the estimate $\widehat{W}=m$. 
If $\vecy \in \mathcal{R}_0$, the decoder declares that the transmitter was idle. 

Following~\cite{Weinberger14-09a}, we next define the \gls{fa}, \gls{md}, and \gls{ie} probabilities as follows: 
\begin{IEEEeqnarray}{lCl}
    P\sub{\gls{fa}} &\triangleq& \sum\limits_{m=1}^M P_{\randvecy|\randvecx=\boldsymbol{\varnothing}}[\randvecy \in \mathcal{R}_m]\label{eq:FA}\\
    P\sub{\gls{md}} &\triangleq& \frac{1}{M}\sum\limits_{m=1}^M P_{\randvecy|\randvecx=\vecc_m}[\randvecy \in \mathcal{R}_0]\label{eq:MD}\\
    P\sub{\gls{ie}} &\triangleq& \frac{1}{M}\sum\limits_{m=1}^M \sum\limits_{\substack{k=0 \\ k\neq m}}^M P_{\randvecy|\randvecx = \vecc_m}[\randvecy \in \mathcal{R}_k].\label{eq:IE}
\end{IEEEeqnarray}
Observe that $P\sub{\gls{ie}}$ comprises the probability of \gls{md} and the probability of decoding a wrong codeword.

An $\lro{M,n,\efa,\emd,\ie}$--code is a code with blocklength $n$, $M$ codewords, and $P\sub{\gls{fa}} \leq \efa$, $P\sub{\gls{md}} \leq \emd$, and $P\sub{\gls{ie}} \leq \ie$. 
Similar to~\cite{Polyanskiy11-08}, it will turn out convenient to consider a more general notion of \emph{randomized code} in which both transmitter and receiver are equipped with a common
randomness, which allows them to time-share between deterministic $\lro{M,n,\efa,\emd,\ie}$--codes. 
As usual, the maximum coding rate is defined as 
\begin{multline}
  R^*\lro{n,\efa,\emd,\ie} \\ \triangleq \sup\lrbo{\frac{\log_2\lro{M}}{n} : \exists \lro{M,n,\efa,\emd,\ie}\text{--code} }.
\end{multline}

\section{Nonasymptotic Bounds}\label{sec:nonasymptotic_bounds}

\subsection{Joint Detection and Decoding}
In Theorem~\ref{thm:achievability} below, we present a joint-detection-and-decoding achievability bound that generalizes
the $\beta\beta$ bound for the case of genie-aided detection presented in~\cite[Th.1]{Yang18}. 
\begin{thm}\label{thm:achievability}
Let $P_{\randvecx}$ be an arbitrary distribution on $\mathcal{X}^n$, and let $P_{\randvecy}$ be the corresponding output distribution.
Fix an arbitrary auxiliary distribution $Q_{\randvecy}$ on
$\mathcal{Y}^n$, and two parameters $\delta^{(1)}\in[0,1]$ and $\delta^{(2)}\in[0,1]$. 
For all $M\leq 2/\delta^{(2)} + 1$, there exists a randomized
$(M,n,\efa,\emd,\ie)$-code involving time sharing between four deterministic codes, that satisfies 
\begin{IEEEeqnarray}{rCl}
  \efa & \leq & \delta^{(1)}\label{eq:efa_th2} \\
  \emd & \leq & \alpha_{\delta^{(1)}}(P_{\randvecy}, P_{\randvecy\given \randvecx = \boldsymbol{\varnothing}}) \label{eq:emd_th2} \\
  \ie & \leq & \alpha_{\delta^{(1)}}(P_{\randvecy}, P_{\randvecy\given \randvecx = \boldsymbol{\varnothing}}) \nonumber\\ 
  &&{} + 1-\alpha_{\frac{M-1}{2} \delta^{(2)}}(P_{\randvecy}, Q_{\randvecy}) + \alpha_{\delta^{(2)}}(P_{\randvecx\randvecy}, P_{\randvecx}Q_{\randvecy}).\nonumber\\ \label{eq:ie_th2}
\end{IEEEeqnarray}
\end{thm}
\begin{IEEEproof}
See Appendix~\ref{app:achievability}.  
\end{IEEEproof}
\begin{rem}
Theorem~\ref{thm:achievability} recovers the $\beta\beta$ achievability bound~\cite[Th.~1] {Yang18} when genie-aided detection is
considered and, hence, $ P\sub{\gls{fa}} =  P\sub{\gls{md}} =0$.
Furthermore, for the case of genie-aided detection, by setting $Q_{\randvecy} = P_{\randvecy}$ and using the definition of the Neyman-Pearson test, one recovers from Theorem~\ref{thm:achievability} the \gls{dt} achievability bound~\cite[Th.17]{polyanskiy10-05a}. 
\end{rem}

Next, we present a converse bound that draws inspiration from the $\beta\beta$ converse bound introduced in~\cite[Th.~15]{polyanskiy14-01}
to characterize the empirical output distribution of good channel codes.
\begin{thm}\label{thm:converse}
Let $Q_{\randvecy}$ be an arbitrary distribution on $\mathcal{Y}^n$. 
  Then, every $\lro{M,n,\efa,\emd, \ie}$--code satisfies
  \begin{IEEEeqnarray}{rCl}
    M &&\leq  \sup\limits_{P_{\randvecx}} \biggl\{ \frac{1-\beta_{1-\efa}\lro{P_{\randvecy|\randvecx=\boldsymbol{\varnothing}}, Q_{\randvecy}}} {\beta_{1-\ie}\lro{P_{\randvecx}P_{\randvecy \given \randvecx}, P_{\randvecx}Q_{\randvecy}}}\nonumber\\
    &&\times \mathbb{1}\lrbo{\beta_{1-\emd}\lro{P_{\randvecy}, Q_{\randvecy}} \leq 1- \beta_{1-\efa}\lro{P_{\randvecy \given \randvecx = \boldsymbol{\varnothing}}, Q_{\randvecy}}}\biggr\}.\nonumber\\\label{eq:converse1}
  \end{IEEEeqnarray}
\end{thm}
\begin{IEEEproof}
    See Appendix~\ref{app:converse}.
\end{IEEEproof}
 \begin{rem}\label{rem:converse}
By upper-bounding by one both the indicator function and the numerator of the first term, we recover the metaconverse bound~\cite[Th.
27]{polyanskiy10-05a}---a bound that in our setup would depend only on $\ie$, and, hence, would not be able to illustrate
the impact of the \gls{fa} and \gls{md} requirements  on the size of the codebook $M$. 

  \end{rem}

\subsection{Preamble-Based Detection}\label{sec:separate}
We next show how to adapt the achievability bound in Theorem~\ref{thm:achievability} and the converse bound in Theorem~\ref{thm:converse}
to the case of separate detection and decoding. Let $\randvecx = [\vecx\sub{p},\randvecx\sub{d}]$, where $\vecx\sub{p}$ is a deterministic
preamble sequence of length $\np$, and $\randvecx\sub{d}\sim P_{\randvecx\sub{d}}$ on $\mathcal{X}^{\mathrm{d}}$ is the vector containing the
remaining $\nd = n-\np$ information-carrying symbols. 
Furthermore, let $\randvecy = [\randvecy\sub{p},\randvecy\sub{d}]$ be the received signal upon transmission of $\randvecx = [\vecx\sub{p},\randvecx\sub{d}]$. 
The decoder uses $\randvecy\sub{p}$ to detect if a codeword is present, and if so, decoding is performed over the remaining $\randvecy\sub{d}$ symbols. 
In this setup, packet detection involves a simple binary hypothesis testing problem. 
Hence, the tradeoff between $\efa$ and $\emd$ can be expressed as 
\begin{IEEEeqnarray}{rCl}
    \emd &=& \alpha_{\efa}(P_{\randvecy\sub{p}\given\randvecx\sub{p} = \vecx\sub{p}}, P_{\randvecy\sub{p}\given \randvecx\sub{p} = \boldsymbol{\varnothing}}). \label{eq:pre_det}
\end{IEEEeqnarray}
A bound on $\ie$ can be obtained by following steps similar to the ones reported in Appendix~\ref{app:achievability}.
Specifically, one can show that for an arbitrary $Q_{\randvecy}$ and  $\delta\in[0,1]$, the following bound holds 
 for all $M\leq 2/\delta + 1$: 
 \begin{multline}
    \ie \leq  \emd + (1-\emd)\\
      \times \lrho{1-\alpha_{\frac{M-1}{2} \delta}(P_{\randvecy\sub{d}}, Q_{\randvecy\sub{d}}) +
      \alpha_{\delta}(P_{\randvecx\sub{d},\randvecy\sub{d}}, P_{\randvecx\sub{d}}Q_{\randvecy\sub{d}})}. 
     \label{eq:ach_pre}
\end{multline}
Here,  $P_{\randvecy\sub{d}}$ denotes the output distribution induced by $P_{\randvecx\sub{d}}$. 

A converse bound on $\ie$ for this scenario can be obtained by a direct application of the metaconverse theorem~\cite[Th.
27]{polyanskiy10-05a}.
Specifically, for every auxiliary distribution $Q_{\randvecy\sub{d}}$ on $\mathcal{Y}^{\nd}$, we have that 
\begin{IEEEeqnarray}{rCl}
M &\leq & \sup\limits_{P_{\randvecx\sub{d}}} \frac{1} {\beta_{1-\ie}\lro{P_{\randvecx\sub{d}}P_{\randvecy\sub{d} \given \randvecx\sub{d}}, P_{\randvecx\sub{d}}Q_{\randvecy\sub{d}}}}.\label{eq:conv_pre}
\end{IEEEeqnarray}

\section{Numerical Results}
\subsection{Bounds for the Binary AWGN Channel}\label{sec:bi-awgn-bounds}
In this section, we discuss how to compute the bounds presented in Section~\ref{sec:nonasymptotic_bounds} for the memoryless discrete-time binary-input \gls{awgn} channel.
For this channel, the input-output relation is given by
\begin{IEEEeqnarray}{rCl}
  Y_k &=& X_k + N_k, \quad k=1,\dots,n. \label{eq:ch_bi-awgn}
\end{IEEEeqnarray}
Here,  $X_k\in\mathcal{X} = \{-\sqrt{\rho}, 0, \sqrt{\rho}\}$, where $X_{k}=0$ indicates that the transmitter is idle (in other words, $\varnothing=0$).
The additive noise $N_k$ follows a  $\mathcal{N}\lro{0, 1}$ distribution; hence, $\rho$ can be thought of as the SNR at the receiver.
It then follows that $P_{Y\given X=x}=\normal(x,1)$.
Throughout this section, we consider as input distribution $P_{\randvecx}$ a product distribution $P_{\randvecx}(\vecx)= \prod_{k=1}^{n}P_{X}(x_{k})$
where {$P_{X}(-\sqrt{\rho})=p$, $P_{X}(\sqrt{\rho})=1-p$}, and $P_{X}(0)=0$, with $p\in [1/2,1]$.
By adjusting $p$, we can enforce one of the two nonzero symbols to occur more often in the codebook, which, as we shall see, is beneficial from a detection perspective.
As auxiliary channel, we choose $Q_{\randvecy}(\vecy)= \prod_{k=1}^{n}P_{Y}(y_{k})$, where $P_{Y} = p\mathcal{N}(-\sqrt{\rho},1) +
(1-p)\mathcal{N}(\sqrt{\rho},1)$ is the output distribution induced by $P_{X}$.
To evaluate the achievability bound in Theorem~\ref{thm:achievability}, we note that, for this choice of input distribution, we have that  $Q_{\randvecy}=P_{\randvecy}$. We then use the Neyman-Pearson
lemma to evaluate the $\alpha$ functions in~\eqref{eq:efa_th2}--\eqref{eq:ie_th2}.

Evaluating the converse bound given in Theorem~\ref{thm:converse} for joint detection and decoding is, however, numerically intractable. 
Indeed,~\eqref{eq:converse1} involves an optimization over all possible input distributions, which cannot be avoided due to the presence of the induced output distribution $P_{\randvecy}$ in~\eqref{eq:converse1}. 
One possible approach to sidestep this issue is to upper-bound by one both the indicator function in~\eqref{eq:converse1} and the numerator of
the first term in~\eqref{eq:converse1}. 

The resulting bound 
  \begin{IEEEeqnarray}{rCl}
    M &&\leq  \sup\limits_{P_{\randvecx}}\frac{1}{\beta_{1-\ie}\lro{P_{\randvecx}P_{\randvecy \given \randvecx}, P_{\randvecx}Q_{\randvecy}}}\label{eq:converse-metaconverse}
  \end{IEEEeqnarray}
which, as already remarked, coincides with the metaconverse bound~\cite[Th. 27]{polyanskiy10-05a}, yields also a converse bound for the
genie-aided case.
Note that this bound can be evaluated numerically if we set $p=1/2$ in $Q_{\randvecy}$.
Indeed, for this choice one can invoke~\cite[Lem.~29]{polyanskiy10-05a}, and replace the $\beta_{1-\ie}\lro{P_{\randvecx}P_{\randvecy \given
    \randvecx}, P_{\randvecx}Q_{\randvecy}}$ in the denominator of~\eqref{eq:converse-metaconverse} with $\beta_{1-\ie}\lro{P_{\randvecy \given
\randvecx=\vecx}, Q_{\randvecy}}$, where $\vecx$ is an arbitrary $n$-dimensional vector with entries in {$\{-\sqrt{\rho},\sqrt{\rho}\}$}.
The resulting bound is then independent of $P_{\randvecx}$, and the maximization can be omitted. 

A different approach to assess, although in a weaker way, the tightness of the achievability bound in Theorem~\ref{thm:achievability} for
our choice of parameters, is to evaluate the converse bound in Theorem~\ref{thm:converse} for the same input distribution $P_{\randvecx}$ chosen to evaluate the achievability bound. 
The resulting converse bound should be interpreted as a \emph{random-coding-ensemble converse}. 
It provides an upper bound on the number of codewords that are compatible with the requirement that the \gls{fa}, the \gls{md}, and the
\gls{ie} probabilities, averaged over all random codebooks whose codewords are drawn independently according to $P_{\randvecx}$, do not exceed $\efa$, $\emd$, and $\ie$, respectively.  

To evaluate the preamble-based bounds~\eqref{eq:pre_det}, \eqref{eq:ach_pre}, and~\eqref{eq:conv_pre}, we set $p=1/2$ and proceed similarly
to the joint-detection-and-decoding case.
This time, the input and auxiliary output distributions are defined on vectors of dimension $\nd$. 

To evaluate numerically the bounds discussed so far via the Neyman-Pearson lemma, one needs to compute the following three likelihood ratios:
\begin{IEEEeqnarray}{rCl}
  \imath(\vecx; \vecy) &\triangleq &\log\lro{\frac{\mathrm{d}P_{\randvecy \given \randvecx = \vecx}}{\mathrm{d}P_{\randvecy}} } \label{eq:info_dens}\\
\jmath(\vecx; \vecy) &\triangleq &\log\lro{\frac{\mathrm{d}P_{\randvecy \given \randvecx = \vecx}}{\mathrm{d}P_{\randvecy\given \randvecx=\boldsymbol{\varnothing}}} } \label{eq:jinfo_dens}\\
r(\vecy) &\triangleq &\log\lro{\frac{\mathrm{d}P_{\randvecy}}{\mathrm{d}P_{\randvecy\given \randvecx =\boldsymbol{\varnothing}}}}.\label{eq:rinfo_dens}
\end{IEEEeqnarray}
For our choice of output distribution $Q_{\randvecy}$, these quantities can be evaluated as follows:
\begin{IEEEeqnarray}{rCl}
  \imath\lro{\vecx, \vecy} 
  &=&
  \sum_{k=1}^n \log\lro{\frac{e^{x_k y_k}}{p e^{-\sqrt{\rho}y_k} + (1-p)e^{\sqrt{\rho}y_k}}} \label{eq:biawgn_RN1}\\
  r\lro{\vecy} 
  &=& -\frac{n \rho}{2} + \sum_{k=1}^n \log\lro{p e^{-\sqrt{\rho}y_k} + (1-p) e^{\sqrt{\rho}y_k}} \label{eq:biawgn_RN2} \IEEEeqnarraynumspace \\
  \jmath\lro{\vecx,\vecy} &=&  -\frac{n \rho}{2} + \sum_{k=1}^n x_k y_k.\label{eq:biawgn_RN3}
\end{IEEEeqnarray}
For example,~\eqref{eq:biawgn_RN3} implies that~\eqref{eq:pre_det} can be expressed in the following parametric form: for every $\gamma \in \reals$, we have that 
\begin{IEEEeqnarray}{rCl}
  \efa &=& P_{\randvecy\sub{p}\given \randvecx = \boldsymbol{\varnothing}}\lrho{\sum_{i=1}^{\np} x_i Y_i \geq\gamma + \frac{\np\rho}{2}}\nonumber\\
   &=& Q\lro{\frac{\gamma+\frac{\np\rho}{2}}{\sqrt{\np\rho}}} \label{eq:pre_det1_bi_2}
\end{IEEEeqnarray}
and
\begin{IEEEeqnarray}{rCl}
 \emd &=& P_{\randvecy\sub{p}\given \randvecx\sub{p} = \vecx\sub{p}}\lrho{\sum_{i=1}^{\np} x_i Y_i<\gamma+\frac{\np\rho}{2}}\nonumber\\
 &=& 1-Q\lro{\frac{\gamma-\frac{\np\rho}{2}}{\sqrt{\np\rho}}}.\label{eq:pre_det2_bi_2}
\end{IEEEeqnarray}
\subsection{Numerical Results for the Binary AWGN Channel}\label{sec:num_subsection}
\begin{figure}[!ht]
\centering
\setlength{\belowcaptionskip}{-13pt}
    \input{./Figures/BIAWGN4_globecom.tex}
    \setlength{\belowcaptionskip}{-12pt}
    \caption{Rate bounds for the binary \gls{awgn} with $\ie=10^{-3}$, $\emd=10^{-4}$, and $\efa=10^{-4}$. 
    } 
\label{fig:biawgn_det}
\end{figure}

In Fig.~\ref{fig:biawgn_det}, we compare the maximum coding rate achievable with joint detection and decoding and with preamble-based detection followed by decoding, with the maximum coding rate achievable with genie-aided detection. 
In our simulations, we set $\ie = 10^{-3}$, $\emd = 10^{-4}$, $\efa = 10^{-4}$, and consider three different SNR values: $\rho\in \{0\dB,3\dB,6\dB\}$. 

The achievability bound for joint detection and decoding given by Theorem~\ref{thm:achievability}, and optimized over the choice of $p$ is depicted with blue dashed lines, whereas we use green dashed lines to indicate the achievability bound for the case of preamble-based detection, obtained by evaluating~\eqref{eq:pre_det1_bi_2}, \eqref{eq:pre_det2_bi_2} and \eqref{eq:ach_pre}.
This bound is optimized over the choice of $\np$.  
Furthermore, we use red dashed lines to indicate an achievability bound for the genie-aided-detection case.
Specifically, we  consider the \gls{dt} bound~\cite[Th. 17]{polyanskiy10-05a}, evaluated for the input distribution described in
Section~\ref{sec:bi-awgn-bounds} and  $p=1/2$.

On the converse side, we use blue solid lines to indicate the ensemble-converse bound discussed in Section~\ref{sec:bi-awgn-bounds}, optimized over the choice of $p$. 
The converse bound for the preamble-based detection, obtained by evaluating~\eqref{eq:pre_det1_bi_2}, \eqref{eq:pre_det2_bi_2}, and \eqref{eq:conv_pre} is indicated by green solid lines.
Finally, we use red solid lines to indicate the metaconverse bound~\eqref{eq:converse-metaconverse} for the genie-aided case.  

We observe that the genie-aided-detection bounds approach the 
joint detection and decoding bounds as the blocklength and the SNR increase. In our numerical examples, this occurs when $n\geq190$ for an SNR value of $0\dB$, when $n\geq70$ for an SNR value of $3\dB$, and for $n\geq20$ for an SNR value of $6\dB$. 
On the contrary, the preamble-based strategy is suboptimal over the entire range of blocklength values considered in the figure, although the performance gap decreases as the SNR increases. 
This is expected, since detection is simplified when the power available for the transmission of the  preamble increases (see~\eqref{eq:pre_det1_bi_2} and~\eqref{eq:pre_det2_bi_2}), in which case a shorter preamble sequence suffices. 
Thus, the penalty due to the transmission of the preamble decreases as the SNR and the blocklength grow. 

\begin{figure}[t]
  \centering
  \setlength{\belowcaptionskip}{-13pt}
      \begin{tikzpicture}
\pgfplotsset{
   scaled y ticks = false,
   height=\fwidth*0.3,
   every tick label/.append style={font=\footnotesize},
   xlabel style={font=\footnotesize},
   ylabel style={font=\footnotesize},
    title style={yshift=-6pt,}
}
\begin{groupplot}[
    group style={
    group name=my plots,
    group size = 1 by 1,
    vertical sep=45pt,
    horizontal sep=45pt,
    }
]
   \nextgroupplot[
         width = 0.5\textwidth,
         xlabel={$n$ [channel uses]},
         ylabel = {$p$},
         grid=none,
         xmin = 10,
         xmax=300,
         ymin=0.5,
         ymax = 1,
         ytick pos=left,
         ytick={0.5,0.6,0.7,0.8,0.9,1},
         xtick pos=bottom,
         grid = major
        ]
        \def \fa {./Figures/Data/BIAWGN_IMP_DET_SNR_0dB_ed_1e-03_efa_1e-04_emd_1e-04.csv}
        \def \fb {./Figures/Data/BIAWGN_IMP_DET_SNR_3dB_ed_1e-03_efa_1e-04_emd_1e-04.csv}
        \addplot [name path = p_snr0, color=blue, line width=\linew] table [y index={15}, x index = {0}, col sep=comma] {\fa};\label{leg:biawgn_p_th2_snr0}


        \addplot [name path = p_snr3,color=red!60!black, line width=\linew, dashed] table [y index={15}, x index = {0}, col sep=comma] {\fb};\label{leg:biawgn_p_th2_snr0}

        \coordinate (pt2) at (axis cs: 65,0.6);
        \coordinate (pt3) at ($(pt2)+ (20pt,10pt)$);
        \draw[<-] (pt2)--(pt3) node at ($(pt3) + (6pt,4pt)$) {\footnotesize SNR$=3$~dB};  
        
        \coordinate (pt2) at (axis cs: 160,0.7);
        \coordinate (pt3) at ($(pt2)+ (20pt,10pt)$);
        \draw[<-] (pt2)--(pt3) node at ($(pt3) + (4pt,4pt)$) {\footnotesize SNR$=0$~dB};

\end{groupplot}


\end{tikzpicture}%
      \setlength{\belowcaptionskip}{-12pt}
      \caption{Optimal value of the parameter $p$ in the achievability bound provided in Theorem~\ref{thm:achievability} for joint detection and decoding. 
      Here, $\ie=10^{-3}$, $\emd=10^{-4}$, and $\efa=10^{-4}$. 
      } 
  \label{fig:p-param}
\end{figure}
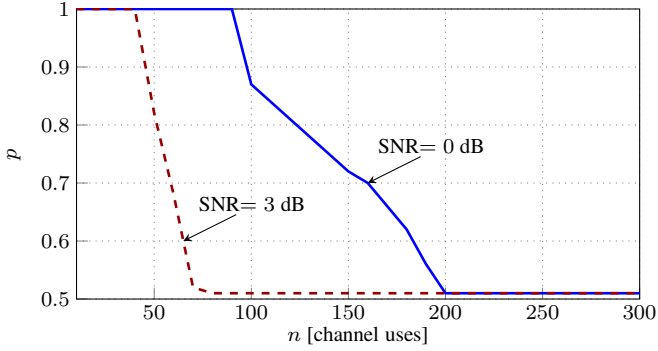

In Fig.~\ref{fig:p-param}, we depict the value of $p$ that maximizes the achievability bound given in Theorem~\ref{thm:achievability} for the case of joint detection and decoding, as a function of the blocklength, for the case $\rho=0\dB$ and $\rho=3\dB$. 
The other parameters are as in~Fig.~\ref{fig:biawgn_det}.
As illustrated in the figure, skewing the input distribution is beneficial for blocklength values where detection is the performance
bottleneck. 
This corresponds to the range of blocklength values for which the joint detection and decoding scheme performs strictly better than the preamble-based scheme, but strictly worse than the genie-aided scheme. 
Note that by setting $p=0.5$, we obtain the capacity achieving input distribution, which maximizes the coding rate in the large blocklength limit. 
However, uniform inputs are not a good choice for detection when packets are short. 
For example, in the scenario depicted in Fig.~\ref{fig:biawgn1a}, by using uniform inputs, the achievable rate with joint detection and decoding would be equal to zero until $n=190$, to then rapidly approach the genie-aided bound for slightly larger blocklength values. 
As shown in the figure, higher rates can be obtained for blocklengths smaller than $190$ by optimizing $p$. 


\section{Conclusion}\label{sec:conclusion}
We have presented nonasymptotic bounds on the largest coding rate achievable in a frame-synchronous communication system where the receiver has to decide on the presence of a packet prior to decoding.
The bounds, which rely on the $\beta\beta$ framework introduced in~\cite{Yang18}, apply to the scenario where detection is performed jointly with decoding over the entire data packet, and to the scenario in which a dedicated preamble is used for detection.

Numerical examples for the case of binary-input AWGN channels indicate that in the short-packet regime, joint detection and decoding yields significant gains in terms of maximum coding rate over preamble-based detection followed by decoding.
Furthermore, there exists a range of blocklength values for which  departing from a uniform input distribution is beneficial, since it facilitates detection.
The tightness of the proposed bounds in the error-exponent regime considered in~\cite{Wang10,Weinberger14-09a,Weinberger17-10a} remains to
be investigated.

\appendices
\section{Proof of Theorem~\ref{thm:achievability}}\label{app:achievability}
Fix a codebook $\{\vecc_{m}\}_{m=1}^{M}$ and let $Z^{(1)}(\vecy)$ and $Z^{(2)}(\vecx,\vecy)$ denote the tests achieving $
\alpha_{\delta^{(1)}}(P_{\randvecy}, P_{\randvecy\given \randvecx = \boldsymbol{\varnothing}}) $ and
$\alpha_{\delta^{(2)}}(P_{\randvecx,\randvecy}, P_{\randvecx}Q_{\randvecy})$, respectively.
For a given received signal $\vecy$, the decoder evaluates  $Z^{(1)}(\vecy)$.
If  $Z^{(1)}(\vecy)=0$, the decoder declares the transmitter to be idle.
If $Z^{(1)}(\vecy)=1$, the decoder declares the transmitter to be active and determines the transmitted message as follows: it computes
$Z^{(2)}(\vecc_m,\vecy)$ for all $m=1,\dots,M$ and returns the smallest index $\tilde{m}$ for which $Z^{(2)}(\vecc_{\tilde{m}},\vecy)=1$.
If no such index is found, it declares a decoding error.
For this decoder, it follows that
  \begin{IEEEeqnarray}{rCl}
    && \efa\lefto(\{\vecc_m\}_{m=1}^{M}\right) = P_{\randvecy \given \randvecx=\boldsymbol{\varnothing}}[Z^{(1)}(\randvecy)=1] \label{eq:efa_th2_proof}\\
    && \emd\lefto(\{\vecc_m\}_{m=1}^{M}\right) = \frac{1}{M} \sum_{m=1}^{M}  P_{\randvecy\given \randvecx=\vecc_{m} }[Z^{(1)}(\randvecy)=0] \label{eq:emd_th2_proof}\\ 
    && \ie\lefto(\{\vecc_m\}_{m=1}^{M}\right) =  \frac{1}{M}\sum\limits_{m=1}^M P_{\randvecy \given \randvecx = \vecc_m}\biggl[ 
        Z^{(1)}(\randvecy)=0 \text{ or } \nonumber \\
    &&\quad  Z^{(2)}(\vecc_{m},\randvecy) =0 \text{ or }   Z^{(2)}(\vecc_{m'},\randvecy) =1, m'<m   
    \biggr]. \IEEEeqnarraynumspace \label{eq:ie_th2_proof}
\end{IEEEeqnarray}
We obtain~\eqref{eq:efa_th2} and~\eqref{eq:emd_th2} by averaging~\eqref{eq:efa_th2_proof} and~\eqref{eq:emd_th2_proof} over all random codebooks $\{\randvecc_m\}_{m=1}^M$ whose codewords are generated independently from $P_{\randvecx}$ and by using the definition of the test $Z^{(1)}(\vecy)$. 
To obtain~\eqref{eq:ie_th2} by averaging~\eqref{eq:ie_th2_proof} over  $\{\randvecc_m\}_{m=1}^M$, we proceed as follows. 
We first apply the union bound on the probability of the union of the three events on the right-hand side of~\eqref{eq:ie_th2_proof} to obtain three probability terms. 
The first term can be evaluated as
\begin{multline}
    \Ex{\{\randvecc_{m} \}_{m=1}^{M} }{ \frac{1}{M} \sum_{m=1}^{M}  P_{\randvecy\given \randvecx=\randvecc_{m} }[Z^{(1)}(\randvecy)=0] } \\=
    \alpha_{\delta^{(1)}}(P_{\randvecy}, P_{\randvecy \given \randvecx=\boldsymbol{\varnothing}}).\label{eq:ach-bound-first-term} 
\end{multline}
Similarly, the second term is given by
\begin{multline}
     \Ex{\{\randvecc_{m} \}_{m=1}^{M} }{ \frac{1}{M} \sum_{m=1}^{M}  P_{\randvecy\given \randvecx=\randvecc_{m} }[Z^{(2)}(\randvecc_{m}, \randvecy)=0] } \\=
     \alpha_{\delta^{(2)}}(P_{\randvecx,\randvecy}, P_{\randvecx}Q_{\randvecy} ).\label{eq:ach-bound-second-term}    
\end{multline}
To evaluate the third term, it is convenient to define a random variable $W$ uniformly distributed over the message set $[1,\dots,M]$, and
to consider the randomized test $Z^{(3)}(\vecy)$, which returns $1$ if $Z^{(2)}(\vecc_{m},\vecy)=1$ for some $m<W$. 
Then
\begin{multline}
\Exop_{\{\randvecc_{m} \}_{m=1}^{M} }\biggl[ \frac{1}{M} \sum_{m=1}^{M}  P_{\randvecy\given \randvecx=\randvecc_{m} }[
Z^{(2)}(\randvecc_{m'},\randvecy) 
     =1,\\ m'<m ]\biggr]
= P_{\randvecy}[Z^{(3)}=1].\label{eq:ach-bound-third-term} 
\end{multline}
Now note that
\begin{IEEEeqnarray}{lCl}
    P_{\randvecy}[Z^{(3}(\randvecy) = 1] &=& 1 - P_{\randvecy}[Z^{(3}(\randvecy) = 0] \nonumber \\
                                         &\leq& 1 - \alpha_{Q_{\randvecy}[Z^{(3)}(\randvecy)=1]}(P_{\randvecy},Q_{\randvecy}).
\end{IEEEeqnarray}
Furthermore, 
\begin{IEEEeqnarray}{lCl}
    Q_{\randvecy}[Z^{(3}(\randvecy) = 1] &\leq& \frac{1}{M}\sum\limits_{j=1}^M (j-1) P_{\randvecx}Q_{\randvecy}[Z^{(2)}(\randvecx,\randvecy) = 1]\nonumber\\
                                         &=& \frac{M-1}{2}\delta^{(2)}.
\end{IEEEeqnarray}
Since the function $\alpha_{\beta}$ is nonincreasing in $\beta$, we conclude that 
\begin{equation}
    P_{\randvecy}[Z^{(3)}(\randvecy) = 1] \leq 1 - \alpha_{\frac{M-1}{2}\delta^{(2)}}(P_{\randvecy},Q_{\randvecy}). \label{eq:bb_end}
\end{equation}
The desired result follows by substituting~\eqref{eq:bb_end} into~\eqref{eq:ach-bound-third-term}, and then by 
combining~\eqref{eq:ach-bound-first-term},~\eqref{eq:ach-bound-second-term}, and~\eqref{eq:ach-bound-third-term}.
To conclude the proof, we proceed similarly to~\cite[Th. 19]{Polyanskiy11-08}, and note that, by Caratheodory’s theorem (see
e.g.~\cite[Th. 15.3.5]{cover91a}), there exists a randomized code that achieves
simultaneously~\eqref{eq:efa_th2}--\eqref{eq:ie_th2} and involves time-sharing between four deterministic codes.

\section{Proof of Theorem~\ref{thm:converse}}\label{app:converse}
Fix a coding scheme and let $P_{\randvecx}$ be the distribution on $\setX^{n}$ induced by the encoder when the messages are uniform.
Let $P_{\randvecy}$ be the corresponding output distribution.
Finally, denote by $\setR_{0},\setR_{1},\dots,\setR_{M}$ the decoding regions.
Note that, by assumption, we have that 
\begin{IEEEeqnarray}{rCl}
    P_{\randvecy}[\randvecy\notin\mathcal{R}_0] &\geq& 1-\emd \label{eq:MD_def2}\\
    P_{\randvecy\given \randvecx = \boldsymbol{\varnothing}}[\randvecy\in\mathcal{R}_0] &\geq& 1-\efa. \label{eq:FA_def2}
\end{IEEEeqnarray}
Let now  $Z(\vecx,\vecy) = \mathbb{1}\{\widehat{W} = W, \vecy \notin \mathcal{R}_0\}$ where  $\widehat{W}$ denotes the estimated message by
the decoder when the transmitted message is $W$. 
Note that 
\begin{equation}
    P_{\randvecx,\randvecy}[Z(\randvecx,\randvecy)=1] =    P_{\randvecx,\randvecy}[\widehat{W} = W] \geq 1-\ie. 
\end{equation}
Furthermore, we also have that 
\begin{equation}
    P_{\randvecx}Q_{\randvecy}[Z(\randvecx,\randvecy)=1] \leq \frac{Q_{\randvecy}[\randvecy \notin \setR_{0} ] }{M}.
\end{equation}
It then follows that 
\begin{equation}\label{eq:sort-of-beta-beta-for-given-code}
    \beta_{1-\ie}(P_{\randvecx,\randvecy}, P_{\randvecx}Q_{\randvecy}) \leq    \frac{Q_{\randvecy}[\randvecy \notin \setR_{0} ] }{M}.   
\end{equation}
Note finally that 
\begin{IEEEeqnarray}{lCl}
    \beta_{1-\emd}(P_{\randvecy},Q_{\randvecy}) &\leq& Q_{\randvecy}[\randvecy\notin \mathcal{R}_0] \nonumber \\
    &=& 1-Q_{\randvecy}[\randvecy \in \mathcal{R}_0]\nonumber\\
    &\leq& 1-\beta_{1-\efa}(P_{\randvecy\given\randvecx=\boldsymbol{\varnothing}},Q_{\randvecy}). \label{eq:rel_FA_MD}
\end{IEEEeqnarray}
We obtain the desired bound by using~\eqref{eq:rel_FA_MD}
in~\eqref{eq:sort-of-beta-beta-for-given-code} to remove the dependence on $\setR_{0}$, and by maximizing over $P_{\randvecx}$, to obtain a bound   
that is valid for every code.

\bibliographystyle{IEEEtran}

\end{document}